\documentclass[reprint,pra,twocolumn,twoside,aps]{revtex4-1}

\usepackage{graphicx}

\usepackage{amsmath, amssymb}
\usepackage{bm}
\usepackage{natbib}
\usepackage{color}
\usepackage{soul,xcolor}

\setstcolor{red}

\begin{document}

\title{Slow light in flight imaging} 
\author{Kali Wilson$^{1,*}$, Bethany Little$^{2}$, Genevieve Gariepy$^1$, Robert Henderson$^3$, John Howell$^2$, Daniele Faccio$^{1}$}
\email{k.wilson@hw.ac.uk, d.faccio@hw.ac.uk\\}
\affiliation{$^1$Institute of Photonics and Quantum Sciences, Heriot-Watt University, Edinburgh EH14 4AS, UK} 
\affiliation{$^2$Department of Physics and Astronomy, University of Rochester, Rochester, New York 14627, USA}
\affiliation{$^3$Institute for Micro and Nano Systems, University of Edinburgh, Alexander Crum Brown Road, Edinburgh EH9 3FF, UK}


\begin{abstract}
Slow-light media are of interest in the context of quantum computing and enhanced measurement of quantum effects, with particular emphasis on using slow-light with single photons.   We use light-in-flight imaging with a single photon avalanche diode camera-array to image \emph{in situ} pulse propagation through a slow light medium consisting of heated rubidium vapour.  Light-in-flight imaging of slow light propagation enables direct visualisation of a series of physical effects including simultaneous observation of spatial pulse compression and temporal pulse dispersion.  Additionally, the single-photon nature of the camera allows for observation of the group velocity of single photons with measured single-photon fractional delays greater than 1 over 1 cm of propagation.  
\end{abstract}

\maketitle

\section{Introduction}
\indent Slow light has been proposed and studied as a resource for building quantum networks; carefully controlled transmission of quantum states of light \cite{Eisaman2005, Chang2014}, tuneable delay lines \cite{Camacho2006, Akopian2011} with fast reconfiguration rates \cite{Camacho2007b} and spectral resolution \cite{Wildmann2015}, and tuneable delay of entangled images \cite{Marino2009} are just a few examples.  More recently slow light has been exploited to enhance measurement, either through improving the sensitivity of interferometers \cite{Shi2007b}, or by enhancing effects that would normally be too small to observe, such as rotary photon drag \cite{Franke-Arnold2011} and Fresnel light dragging \cite{Safari2016, Kuan2016}, or the proposed observation of gravitational deflection of classical light \cite{Dressel2009}. \\
\indent In the context of quantum optics and communications, there is significant interest in the use of slow light as an optical buffer where signals can be stored and resynchronised.
Large fractional delays for single photons are essential for such quantum buffers and memories. Slow group velocities ($v_g \sim 0.003 c$) for single photons have been demonstrated previously \cite{Eisaman2005} in an electro-magnetically induced transparency (EIT) configuration with narrow pulse bandwidths (on the order of MHz), required to fit within the EIT window, and correspondingly large temporal pulse widths on the order of 150 ns. Such long pulse widths limit the overall fractional pulse delay to FD = $\Delta t / \tau < 1$, where $\Delta t$ is the net pulse delay  and $\tau$  is the temporal pulse width \cite{Boyd2005}.  Progress has been made in the high pulse intensity (many photon) regime \cite{Novikova2007, Chen2013}, with an $FD \sim 72$ reported by Chen \emph{et. al.} \cite{Chen2013}. However such EIT schemes are still limited to narrow bandwidth pulses and have not yet been applied to single photons.  \\
\indent Another promising slow light scheme employs an absorption doublet in a hot atomic vapour, such as rubidium or cesium.  In this scenario the frequency of the light is tuned in the region between two absorption resonances with no need for an additional pump beam as required for EIT.  Moreover, large bandwidths on the order of 1-10 GHz are available so that short pulses and large fractional pulse delays $FD > 10$ are possible both for classical pulses \cite{Camacho2007b} and for single photons \cite{Akopian2011}. \\  
\indent In this work, we take advantage of light-in-flight imaging techniques in conjunction with a camera comprised of an array of single-photon avalanche diodes (SPAD camera) \cite{Gariepy2015, Richardson2009a, Richardson2009b} to simultaneously detect \emph{in situ} both the spatial compression and the temporal  dispersion of a pulse of light traveling through a slow-light medium.  Here the medium consists of a hyperfine absorption doublet in hot Rb vapour.  Generally slow light effects have been characterised as the net effect of a pulse propagating through the slow-light medium, i.e. as a pulse delay time measured with a fast photodiode at the output of the medium \cite{Camacho2006}. \emph{In situ} imaging of slow light therefore provides a new approach for studying the physics of such media and enables observation of pulse propagation dynamics, as well as single-photon dynamics as a result of the single photon sensitivity of the camera.  Indeed, we observe a significant delay, on the order of nanoseconds, in the detection of the  photons scattered as the pulse enters the slow-light medium.  This  lag in scattered-photon arrival time is a direct visualisation of the slowing down of the single-photon group velocity. The pulses used here had a temporal full width at half maximum (FWHM) of $\tau \sim 1$ ns, corresponding to a frequency FWHM of $\Delta \nu \sim 440$ MHz, with measured group velocities as low as $v_g \sim 0.006 c$.  At these low group velocities we observe a full fractional pulse delay of up to $FD \sim 40$ over 7 cm of propagation, and $FD \sim 5$ for the scattered single photons, which propagate through $\sim 1$ cm of Rb vapour prior to exiting the cell en-route to the camera.     \\
\section{Slow light medium}
\indent We work in the slow-light region of the dispersion curve in between the $^{85}$Rb $D_1$ hyperfine transitions. Following Camacho \emph{et al.} \cite{Camacho2006}, we approximate the real $n^\prime$ and imaginary $n^{\prime \prime}$ components of the index of refraction $n = n^\prime + i n^{\prime \prime}$ to be
\begin{align}
n^{\prime} &   \approx 1 + \frac{A \delta}{{\omega_0}^2} + \frac{3 A \delta^3}{{\omega_0}^4}, \label{eqn:n} \\
n^{\prime \prime} & \approx \frac{A \gamma}{{\omega_0}^2} + \frac{3 A \gamma \delta^2}{{\omega_0}^4}.
\end{align}
Here $A$ is a constant that characterises the strength of the susceptibility, $\omega_0 = (\omega_2 - \omega_1)/2 \sim 2\pi$ x $1.5$ GHz is half the separation between the two hyperfine transitions comprising the doublet, $\delta$ is the detuning from the midpoint between the two transitions, and $\gamma = \pi \times 5.75$ MHz is the half width at half maximum of the transitions.  For the above approximation to be valid we assume small detuning $|\delta| << \omega_0$, relatively large pulse bandwidths $1/\tau > \gamma$, and that the parameters of the medium satisfy the condition $\omega_0 >> \gamma$, with a susceptibility $\chi << 1$.   Within these reasonable assumptions, and evaluating the absorption coefficient $\alpha = 2\omega n^{\prime \prime} / c$ in between the two resonances we have
\begin{equation}
\alpha_{\delta = 0} = \frac{2\omega A \gamma}{c {\omega_0}^2}, \label{eqn:beta}
\end{equation}
where $\omega = (\omega_1 + \omega_2)/2 + \delta$, and $c$ is the vacuum speed of light. We can then define the group velocity
\begin{equation}
v_g \approx \frac{c}{\omega (dn^\prime / d\delta)} \approx \frac{2\gamma}{\alpha_{\delta = 0}}, 
\end{equation}
and write the group velocity in terms of the optical depth $OD = \alpha L$ of the Rb vapour
\begin{equation}
v_g \approx \frac{2\gamma L}{OD_{\delta = 0}}.
\end{equation}
Since the optical depth depends on temperature, we tune the group velocity by varying the temperature of the Rb vapour, with hotter temperature corresponding to slower group velocity.\\
\begin{figure}[h!]
\includegraphics[width=8cm]{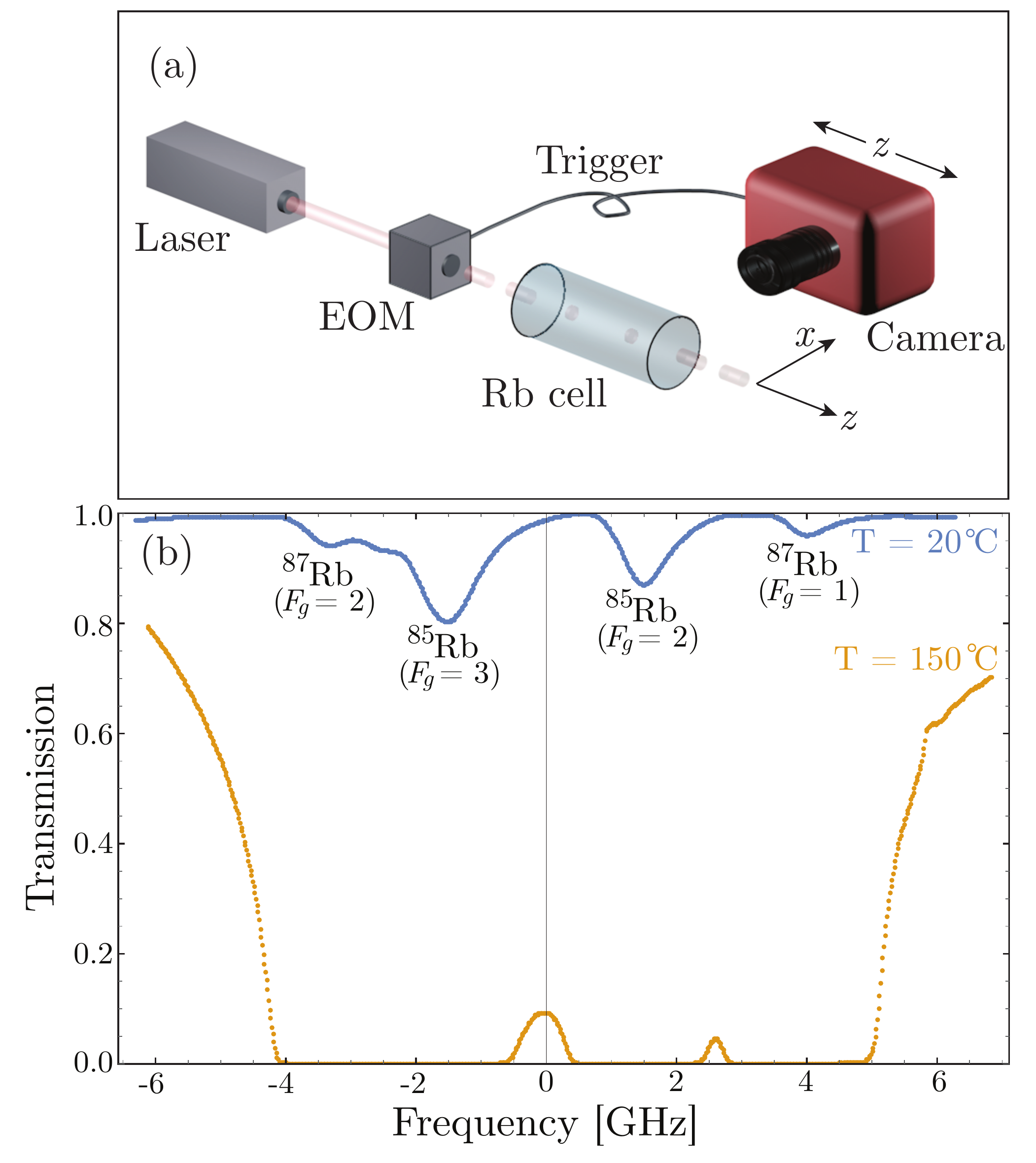}
 \caption{(a) Experimental setup (not to scale). The EOM chops the laser beam into pulses, which propagate through the Rb vapour cell, from left to right along the $z$ axis.  The SPAD camera is set up to capture light scattered by the Rb vapour along the $x$ axis.  The camera has a 5-cm field of view (resolution of $\delta z = 1.6$ mm) and is scanned along the $z$ axis by 4-cm increments. (b) Transmission spectra for the $L =  7.18$ cm cell.  A temperature of $T \sim 150 ^{\circ}$C corresponds to an $OD \sim 2.5$ at $\delta \sim 0$ GHz, with a group velocity $v_g = c/183$, determined from the SPAD data.  The room temperature transmission spectrum, $T \sim 20 ^{\circ}$C, is provided for reference. }
  \label{fig:Exp}
\end{figure}
\section{Experimental layout}
\indent Our experimental slow light setup is shown in Fig.~\ref{fig:Exp}(a). We use a Photodigm distributed Bragg reflector laser diode with wavelength $\lambda = 795$ nm tuned to the middle of the $^{85}$Rb $D_1$ hyperfine doublet, corresponding to $\delta \sim 0$ GHz as shown in the representative transmission spectra in Fig.~\ref{fig:Exp}(b) The laser wavelength is tuned via current and temperature, and is sufficiently stable without the need for servo-locking.  We use an electro-optical modulator (EOM) operating at 25-MHz with a 2\% duty cycle to modulate the laser intensity, creating nearly Gaussian pulses with a temporal FWHM of $\tau \sim 1$ ns, corresponding to a frequency FWHM of $\Delta \nu \sim 440$ MHz, and pulse energy on the order of 1 pJ.  The pulses travel through a heated glass cell containing Rb vapour as shown in Fig.~\ref{fig:Exp}(a). We use two different cells, with outer lengths of $L = 7.18$ cm and $L = 30$ cm, respectively. Figure~\ref{fig:Exp}(b) shows a representative transmission spectrum with the 7.18-cm cell heated to $T \sim 150 ^{\circ}$C, and $OD \sim 2.5$ at $\delta \sim 0$ GHz (orange trace). The transmission spectrum for a room-temperature gas, $T \sim 20 ^{\circ}$C is given for comparison (blue trace). \\
\section{Light in flight imaging}
\indent Following the method of Gariepy \emph{et al.} \cite{Gariepy2015}, we operate a camera consisting of a 32x32 array of single-photon avalanche diodes (SPADs) in the time-correlated, single-photon counting mode, with the SPADs triggered off of the EOM.  We detect photons scattered by the slow-light medium that travel out of the medium perpendicular to the direction of pulse propagation, along the $x$ axis as shown in Fig.~\ref{fig:Exp}(a).  The scattered light is imaged onto the SPAD camera  with a standard camera lens (Sigma 18-35 mm F1.8 DC HSM) with an $f$-number of 1.8 and a focal length of 18 cm, resulting in a 5-cm field of view in the plane of the pulse propagation.   We translate the camera along the $z$ pulse propagation axis by increments of $\Delta z_{\mathrm{cam}} = 4$ cm so as to scan the pulse propagation along the entire length of the Rb cell. The full pulse propagation is then stitched together from the set of  scans.  We also acquired a reference scan for a pulse propagating in air at $v_g \sim c$; we used smoke to provide extra scatterers due to the low pulse energy ($\sim 1$ pJ).  A primary advantage of the SPAD camera  is that it gives us both spatial information with resolution on the order of $\delta z =  1.6$ mm, and temporal information with resolution on the order of $\delta t = 110$ ps set by the impulse-response function of the SPADs.   For each measurement, we typically collect 100,000 frames, each with an exposure time of $500~\mu$s, resulting in a histogram of photon arrival times associated with each pixel as shown in Fig.~\ref{fig:raw}(b) and (c).  These parameters result in $\sim 7$ x $10^{-5}$ detected photons per pulse per pixel, well within the photon-starved regime, with $\sim5$ x $10^{-6}$  detected photons per pulse per pixel corresponding to the signal from the nanosecond pulse and the rest due to continuous-wave (CW) leakage through the EOM. \\
\indent Our data processing steps are as follows.  First, we subtract a background scan to account for any background photons due to light sources other than the laser. Second, we fit the histogram of photon arrival times associated with each pixel of the SPAD array with a Gaussian plus a constant offset that accounts for CW leakage from the EOM. We then subtract off the CW-leakage offset, and filter by the Gaussian FWHM $\tau$ and the signal-to-noise ratio (SNR), i.e. we set pixels to zero if they have low SNR or $\tau$ out of range.  Lastly, for the data shown in Figs. 2 and 3, we replace each temporal histogram with its best fit Gaussian.   \\
 \begin{figure}[h!]
\centering
\includegraphics[width=8cm]{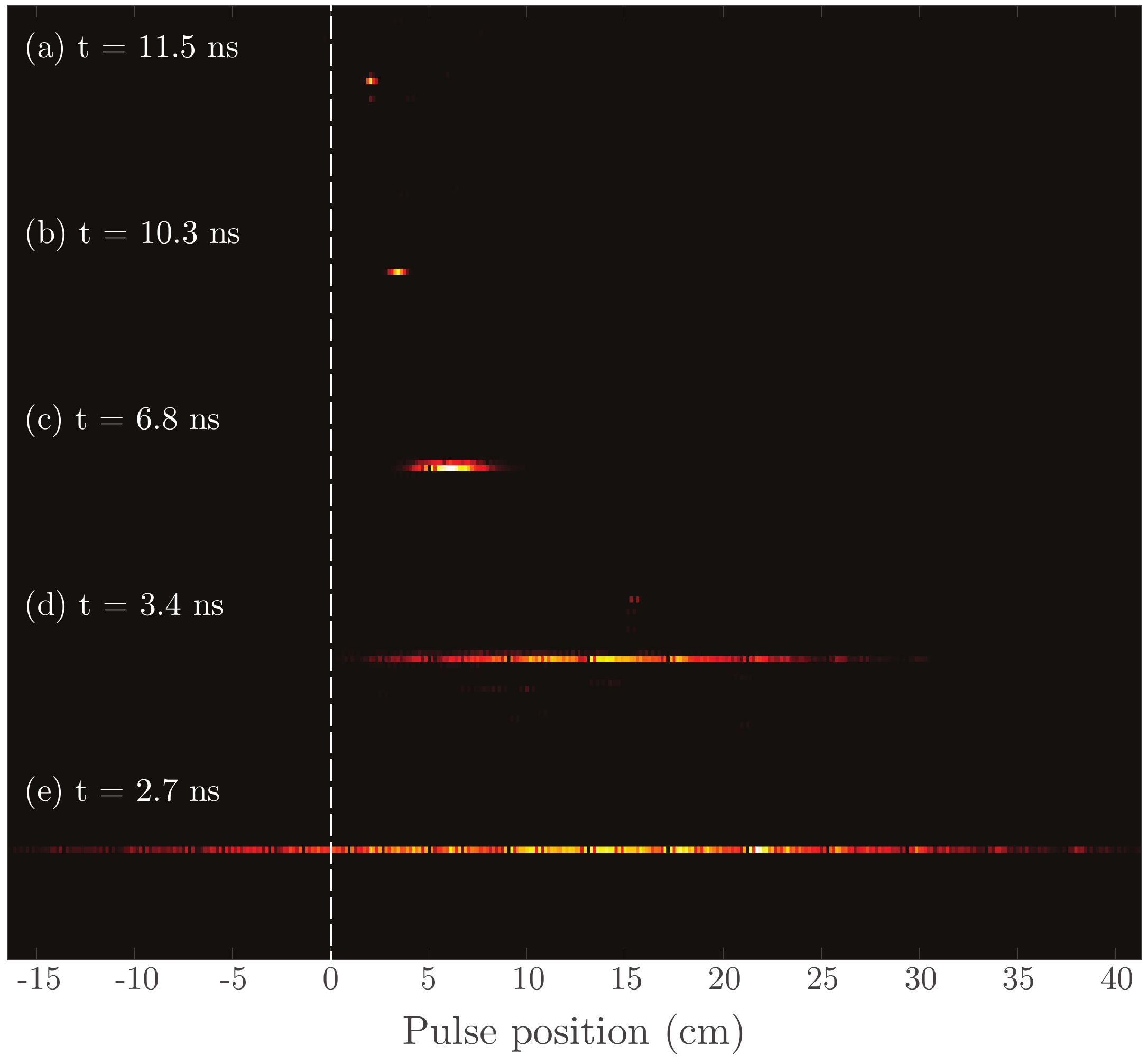}
 \caption{Pulse compression for varying group velocity.  The real time $t$ is given for each pulse, with $t = 0$ ns corresponding to the time when the pulse enters the Rb vapour at $z = 0$ cm. The white dashed line denotes the position at which the light enters the Rb vapour. (a)-(d) correspond to group velocities of $v_g = c/183,~   c/81,~   c/24,$  and  $c/2.6$, with temperatures of $T \sim 150, 130, 106,$ and $50~^{\circ}$C, respectively. (e) shows the pulse propagating through air to provide a reference for light traveling at $v_g \sim c$.    See corresponding video (supplemental information) for the full pulse propagation.  A 7-cm cell was used to obtain the higher temperatures necessary for (a) and (b), while a 30-cm cell was used for (c) and (d). }
 \label{fig:compress}
 \end{figure}
  \begin{figure}[h!]
\centering
\includegraphics[width=7cm]{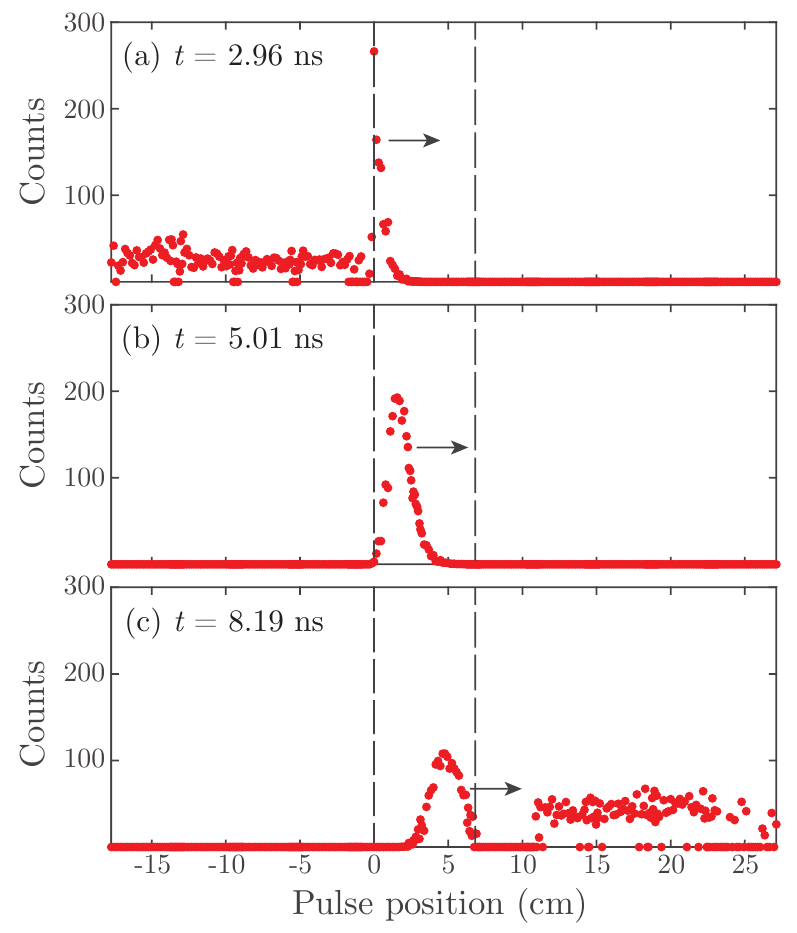}
 \caption{Pulse compression.  Subsequent frames show the pulse (a) propagating through air, (b) entering the heated Rb vapour cell ($L \sim 7$ cm) and compressing spatially due to the decrease in group velocity, and (c) exiting the cell and decompressing. The arrow indicates the direction of pulse propagation. See corresponding video (supplemental information) for the full pulse propagation.  Pulse amplitudes are scaled with respect to the initial pulse amplitude (initial camera postion) to account for variation in pulse energy between scans. }
 \label{fig:interface}
 \end{figure}
 \section{Pulse compression}
\indent A dramatic application of light-in-flight imaging is the \emph{in situ} observation of pulse compression and slowing in the slow-light medium.  Previous observations of slow light in Rb vapour have been limited to observing the net pulse delay at the output of the medium with respect to a reference pulse traveling through air. In contrast, light-in-flight imaging allows us to track the pulse propagation directly \emph{inside} the medium, and to observe the spatial pulse compression as shown in Fig.~\ref{fig:compress}.   Ignoring dispersion (discussed in more detail in Sec.~\ref{sec:disp}), the pulse temporal FWHM is independent of its group velocity, so the spatial extent of the pulse $\Delta z = v_g \tau$ must decrease as $v_g$ decreases.  Figures~\ref{fig:compress}(a)-(d) show the spatial extent of a pulse propagating with group velocities of $v_g = c/183, ~c/81, ~c/24$ and $c/2.6$, respectively, while Fig.~\ref{fig:compress}(e) shows pulse propagation through air for comparison. All pulses are propagating from left to right.  For a pulse propagating in air,  $\tau = 1$ ns corresponds to $\Delta z = 30$ cm.  Therefore, a group velocity of $v_g = c/183$ results in a pulse compressed to $\Delta z \sim 1.6$ mm. The time counters shown for each pulse correspond to the real time, with $t = 0$ ns corresponding to the time when the front edge of the pulse enters the Rb vapour at $z = 0$ cm, indicated by the dashed white line  in Fig.~\ref{fig:compress}. A full video of the pulse propagation for the range of $v_g$ is provided in the supplemental information. \\
\indent Figure~\ref{fig:interface} highlights the pulse compression as the light enters the Rb vapour.  When a pulse enters the slow-light medium the front edge of the pulse slows down, while the trailing end keeps moving forward at the speed of light.  This results in a pulse pile up at the interface.  Figure~\ref{fig:interface}(a) shows the leading edge of the pulse entering the medium. Fig.~\ref{fig:interface}(b) shows the compressed pulse fully contained within the medium ($L \sim 7$ cm), and Fig.~\ref{fig:interface}(c) shows the pulse exiting the cell and decompressing.  See the corresponding video for the full pulse propagation (supplemental information). \\
\begin{figure}[h!]
\includegraphics[width=0.48\textwidth]{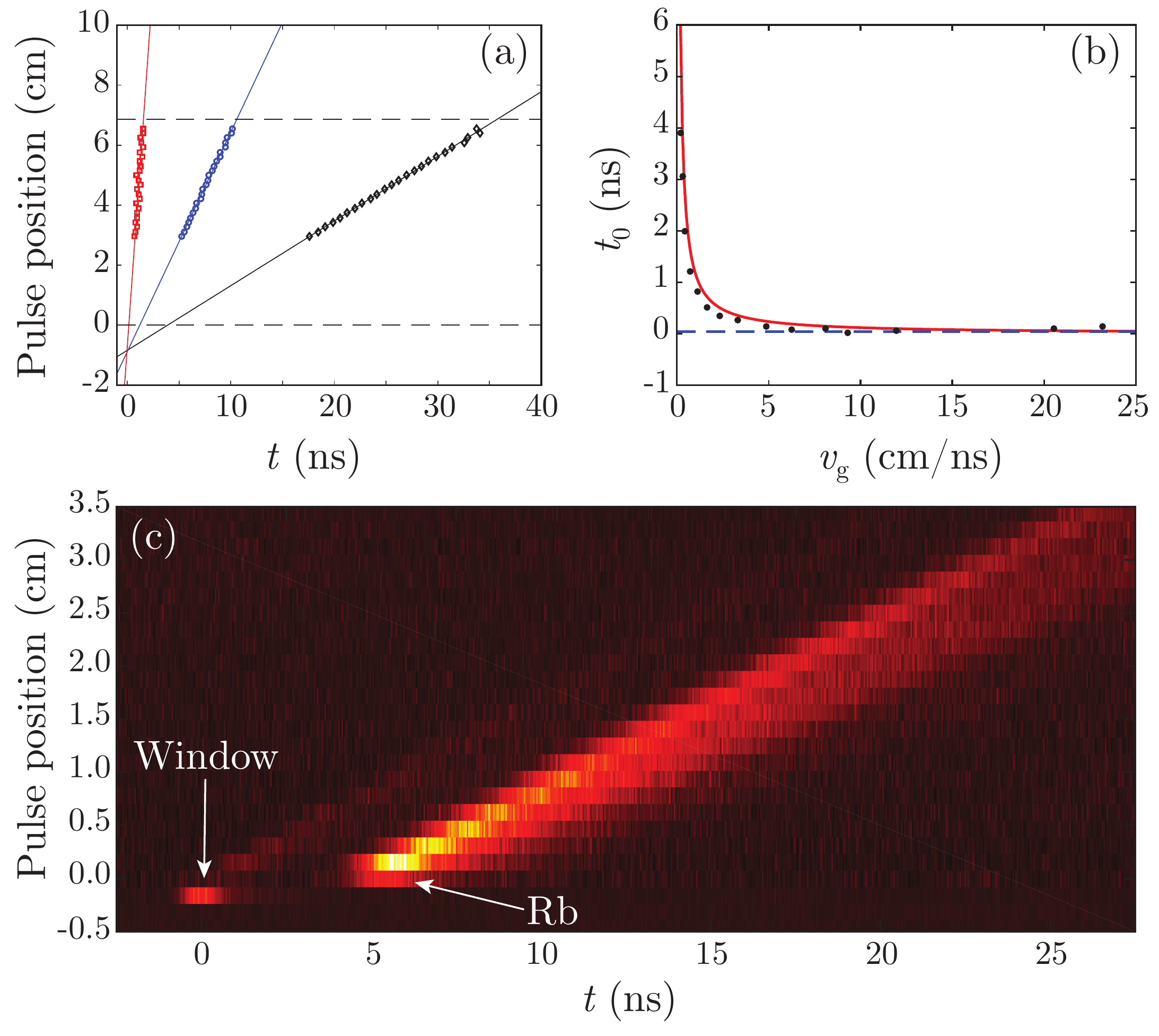}
   \caption{(a) Representative plots of pulse position along the beam path $z$ versus the scattered photon arrival time $t$, shown for group velocities $v_g \sim c/6$ (red squares), $v_g \sim c/41$ (blue circles) and $v_g \sim c/139$ (black diamonds).  Horizontal dashed lines denote the inner edges of the Rb vapour cell, where $z = 0$ corresponds to where the pulse enters the Rb vapour.   Solid lines are fits $z = v_g(t-t_0)$ to the data.  (b)  Photon-lag time versus group velocity.  Black circles: extrapolated time $t_0$ at which the light entered the Rb vapour versus $v_g$.  Red solid line: calculated time lag $t_{0} = r_\mathrm{cell}/v_g$, where $r_\mathrm{cell} = 1.175$ cm is the inner radius of the vapour cell.  Blue dashed line: time lag $t_{0} = r_\mathrm{cell}/c = 0.04$ ns that one would expect if the scattered photons traveled out of the medium at $c$.   (c) Raw temporal profiles corresponding to each pulse propagation position for $v_g \sim c/183$. Note that the light scattered just after the pulse enters the Rb vapour arrives at the camera  approximately 5.3 ns later than the light scattered by the glass window (indicated by white arrows). }
     \label{fig:zvt}
\end{figure}
\section{Single photon lag time}
\indent In the photon-starved regime, the SPAD camera enables the observation of the \emph{single} photon group velocity as demonstrated by Fig.~\ref{fig:zvt}. Figures~\ref{fig:zvt}(a) and \ref{fig:zvt}(b) correspond to a set of scans taken with the camera at a single $z$ position, and falling Rb temperature, i.e., decreasing optical depth and increasing group velocity \cite{note1}. Figure~\ref{fig:zvt}(a) shows the pulse position along the beam path $z$ plotted versus the scattered-photon arrival time $t$ for three representative group velocities $v_g \sim c/139$ (black diamonds), $c/41$ (blue circles), and $c/6$ (red squares). Here $t$ corresponds to the temporal location of the centre of the histogram of scattered-photon arrival times recorded by a SPAD pixel located at position $z$.  The pulse arrival times are determined as the peak position of the best Gaussian fit to the temporal histograms for each $z$ position. Horizontal dashed lines correspond to the inner edges of the Rb vapour cell, where $z = 0$ corresponds to the position where the pulse enters the Rb vapour.   Solid lines are fits $z = v_g(t-t_0)$ to the data, where $v_g$ is the group velocity and $t_0$ is the time at which the pulse enters the cell.  Note the time lag $t_0 \neq 0$, which corresponds to the time it takes the photons that are scattered by the medium to travel out of the cell ($\Delta x \sim r_\mathrm{cell} \sim1$ cm) as they propagate to the camera.  Given the single-photon nature of the SPAD camera detection, the increase in scattered-photon arrival time $t_0$ with decreasing group velocity demonstrates that the slowing down of the group velocity observed for the main pulse propagating along the cell, also applies to the single photons detected by the camera. \\
\indent In Fig.~\ref{fig:zvt}(b), we plot the time lag $t_0$ versus measured group velocity $v_g$ (black circles) for the full set of group velocities.  For comparison we plot the anticipated time lag $t_{0} = r_\mathrm{cell}/v_g$ (red solid line) assuming the pulse propagates down the centre of the vapour cell, where $r_\mathrm{cell} = 1.175$ cm is the inner radius of the vapour cell.  The two curves are in good qualitative agreement and clearly in disagreement with the time lag $t_{0} = r_\mathrm{cell}/c = 0.04$ ns (blue dashed line) that one would expect if the scattered photons traveled out of the medium at the vacuum speed of light $c$.  Figure~\ref{fig:zvt}(c) shows the difference in arrival time for photons scattered in the glass window at the entrance facet of the cell and the arrival time for photons scattered by the Rb vapour for pulse propagation at $v_g \sim c/183$. The photons scattered just after the pulse enters the Rb vapour arrive at the camera approximately 5.3 ns later that the photons scattered by the glass entrance window. Considering that the input pulse is 1 ns long, this last measurement provides direct visualisation of a fractional delay for single photons of $FD = \Delta\tau/\tau\sim 5$ over a propagation distance $\Delta x \sim 1$ cm, which is significantly higher than what can be achieved for single-photons in EIT systems \cite{Eisaman2005}, and in keeping with a recent demonstration in a double-resonant configuration similar to that used here \cite{Akopian2011}.  \\
\begin{figure}[h!]
\centering
\includegraphics[width=0.48\textwidth]{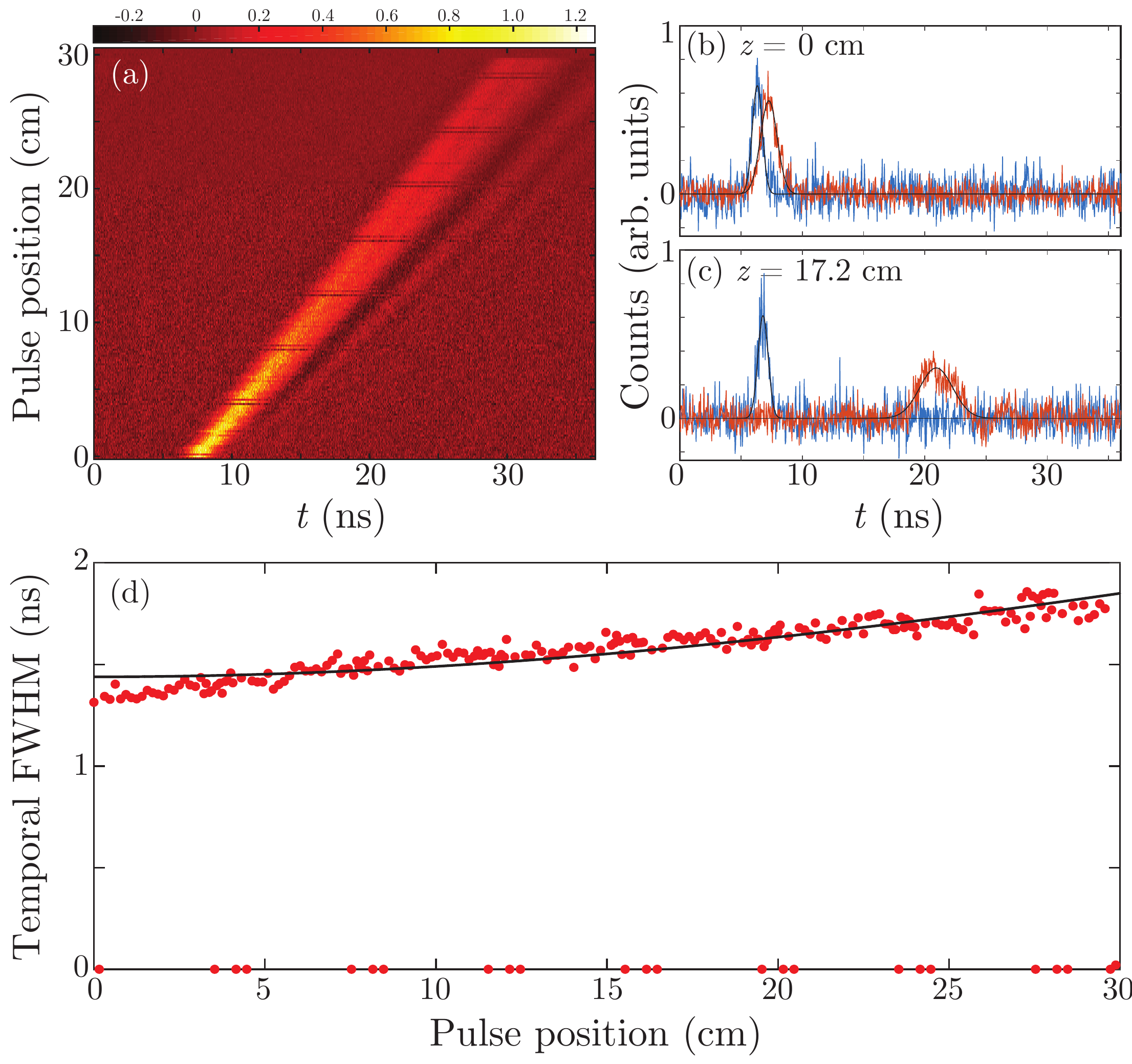}
 \caption{Pulse dispersion observed for a pulse propagating through a Rb vapour cell of length $L = 30$ cm, with $v_g \sim c/24$ ($T \sim 104 ^\circ$C). (a) Plot of raw pulse temporal profiles (horizontal axis) for increasing propagation distance $z$ (pulse position, vertical axis).  (b)-(c) Raw histograms and Gaussian fits to the temporal pulse profiles for $z = 0$ cm and $z =  17.2$ cm, respectively ($z = 0$ cm corresponds to the temporal pulse profile directly after it enters the Rb vapour).  Blue traces correspond to a pulse in air for comparison. Relative amplitudes have been scaled (air: $N_\mathrm{counts}$/70, Rb: $N_\mathrm{counts}$/200) to allow for a comparison of the pulse shape.  (d) Representative plot of pulse temporal FWHM $\tau$ versus pulse position $z$ (red circles) for $T \sim 100~^\circ$C, OD $\sim 0.63$ and $v_g \sim c/13$. Solid black line: data fit with the function $\tau(z) = \sqrt{{\tau_0}^2 + (Bz/\tau_0^2)^2} $, where $\tau_0 = 1.44$ ns is the pulse FWHM width at $z = 0$ and $\beta_3 = B/\sqrt{16 (\ln(2))^{3}}$ = 0.035 ns$^3$/cm. Noisy pixels, corresponding to the repeating line artefacts observed in (a), are set to $\sigma = 0$ and excluded from the fit shown in (d). }
 \label{fig:raw}
 \end{figure}
 \section{Dispersion} \label{sec:disp}
\indent Finally, we note that the light-in-flight technique also measures the pulse dispersion directly as a function of pulse propagation distance.  Figure \ref{fig:raw} shows pulse broadening and distortion for a $\tau \sim 1$ ns pulse traveling through the 30-cm long Rb vapour cell, with $OD \sim 0.92$ (Rb $T \sim 104~^\circ$C) and $v_g \sim c/24$.  The orange traces in Figs.~\ref{fig:raw}(b) and ~\ref{fig:raw}(c) show temporal profile line-outs taken from Fig.~\ref{fig:raw}(a) at $z = 0$ cm and $z = 17.2$ cm, respectively. The blue traces show the temporal profile recorded for pulse propagation in air, and the relative amplitudes of the two traces have been scaled to allow for a comparison of pulse shape. \\
\indent Pulse broadening due to pure quadratic dispersion should follow the form 
\begin{equation}
\tau(z)^2 = \tau_0^2 + \left( \frac{4 \ln(2) ~\beta_2 ~z}{\tau_0} \right)^2  + \left( \frac{4 (\ln (2))^{3/2} ~\beta_3 ~z}{\tau_0^2} \right)^2
 \label{eq:disp}
\end{equation} 
where $\tau_0$ is the FWHM pulse width at $z = 0$ and 
\begin{equation}
\beta_i = \frac{1}{c} \frac{d^i (\omega n^\prime(\omega)) }{d\omega^i} \bigg|_{\omega = (\omega_1 + \omega_2)/2}
\end{equation}
where $\beta_2$ is the second order contribution to the group velocity dispersion (GVD) and $\beta_3$ is the third-order dispersion coefficient \cite{Agrawal2001}.  For the atomic media in question the third-order dispersion term dominates and we can ignore the effects of the second-order GVD. Here the third-order dispersion contributes to both pulse broadening, i.e., GVD, and pulse distortion seen as intensity ripples on the leading edge of the temporal pulse profiles shown in Fig.~\ref{fig:raw}(a) \cite{Camacho2007b}.  Figure~\ref{fig:raw}(d) shows an example of the evolution of the temporal FWHM of the pulse as it propagates along $z$,  for slow-light media with $v_g \sim c/13$ (Rb $T \sim 100^{\circ}$C, $OD \sim 0.63$). The solid line is obtained by fitting the data to Eq.~\eqref{eq:disp}, setting $\beta_2 = 0$ and allowing $\tau_0$ and $\beta_3$ to be parameters of the fit.  The measured values of $\beta_3$ for the three decreasing group velocities $v_g = c/2.6$, $c/13$, and $c/24$ are $\beta_3 = 0.013$, $0.035$, and $0.238$ ns$^3$/cm, respectively, and increase as expected with increasing pulse slowing. We note that the quality of the fit decreases for slower group velocities which we attribute to contributions from detuning dependent absorption \cite{Camacho2007b}.\\
\section{Conclusion}
\indent Single-photon light-in-flight photography allows direct visualisation of the propagation of slow light in a hot atomic vapour. The technique provides direct evidence of pulse compression in space and of single photon delays with fractional delays greater than 1 over $\sim 1$ cm of propagation. In particular, the absorption-doublet slow-light medium used here confirms that extremely large fractional delays are readily achievable with relatively large pulse bandwidths or short nanosecond-pulse durations, similar, for example, to those obtained from quantum-dot single-photon sources \cite{Akopian2011}. The same technique also reveals finer details such as pulse broadening in time due to the large dispersion associated with slow light propagation. The results on the one hand highlight the utility of single-photon light-in-flight measurements for fundamental studies in slow-light media and on the other, provide further support for the use of hot atomic vapours in future single-photon experiments and applications.  

\begin{acknowledgments}
D.F. acknowledges financial  support from the European Research Council under the European
Union's Seventh Framework Programme (FP/2007-2013)/ERC GA 306559, the
Engineering and Physical Sciences Research Council (EPSRC, UK, grants EP/M006514/1,
EP/M01326X/1).
\end{acknowledgments}


\begin{thebibliography}{20}
\bibitem [{not()}]{note1}%
  \BibitemOpen
  \href@noop {} {}\bibinfo {note} {For the data shown in Fig. 4(a) and (b) the
  EOM pulse (25-MHz, 8\% duty cycle) was fed into a tapered amplifier resulting
  in pulse energies $\sim 30$ nJ, with $\tau \sim 3.2$ ns. We used an exposure
  time of 5 $\mu$s resulting in $\sim5$ x $10^{-4}$ detected photons per pulse
  per pixel corresponding to the pulse signal and $\sim3$ x $10^{-3}$ photons
  per pulse per pixel due to CW leakage through the EOM.}\BibitemShut {Stop}%
\end{thebibliography}
%

\end{document}